\documentclass[fleqn,usenatbib]{mnras}

\usepackage{newtxtext,newtxmath}
\usepackage[T1]{fontenc}
\usepackage[utf8]{inputenc}

\usepackage[usenames,dvipsnames,svgnames,table]{xcolor}
\usepackage{comment}
\usepackage{subcaption}
\usepackage{multirow}
\usepackage[flushleft]{threeparttable}

\DeclareRobustCommand{\VAN}[3]{#2}
\let\VANthebibliography\thebibliography
\def\thebibliography{\DeclareRobustCommand{\VAN}[3]{##3}\VANthebibliography}

\newsavebox{\measurebox}
\usepackage{graphicx}	% Including figure files
\usepackage{amsmath}	% Advanced maths commands
\usepackage{scalerel,tikz}
\usetikzlibrary{svg.path}
\definecolor{orcidlogocol}{HTML}{A6CE39}
\tikzset{orcidlogo/.pic={
		\fill[orcidlogocol] svg{M256,128c0,70.7-57.3,128-128,128C57.3,256,0,198.7,0,128C0,57.3,57.3,0,128,0C198.7,0,256,57.3,256,128z};
		\fill[white] svg{M86.3,186.2H70.9V79.1h15.4v48.4V186.2z}
		svg{M108.9,79.1h41.6c39.6,0,57,28.3,57,53.6c0,27.5-21.5,53.6-56.8,53.6h-41.8V79.1z M124.3,172.4h24.5c34.9,0,42.9-26.5,42.9-39.7c0-21.5-13.7-39.7-43.7-39.7h-23.7V172.4z}
		svg{M88.7,56.8c0,5.5-4.5,10.1-10.1,10.1c-5.6,0-10.1-4.6-10.1-10.1c0-5.6,4.5-10.1,10.1-10.1C84.2,46.7,88.7,51.3,88.7,56.8z};
}}
\newcommand\orcidicon[1]{\href{https://orcid.org/#1}{\mbox{\scalerel*{
				\begin{tikzpicture}[yscale=-1,transform shape]
					\pic{orcidlogo};
				\end{tikzpicture}
			}{|}}}}

\usepackage{hyperref}

\title[Looking for LAEs at z=13]{Reionization morphology and intrinsic velocity offsets allow transmission of Lyman-$\alpha$ emission from JADES-GS-z13-1-LA.}

\author[Qin et al.]{Yuxiang Qin~\orcidicon{0000-0002-4314-1810}$^{1}$\thanks{E-mail: Yuxiang.L.Qin@gmail.com}
	and J. Stuart B. Wyithe~\orcidicon{0000-0001-7956-9758}$^{1}$
\\
$^{1}$Research School of Astronomy and Astrophysics, Australian National University, Canberra, ACT 2611, Australia
}

\date{}%Accepted XXX. Received YYY; in original form ZZZ}

\pubyear{2024}

\begin{document}
\label{firstpage}
\pagerange{\pageref{firstpage}--\pageref{lastpage}}
\maketitle

\begin{abstract}
We investigate the detectability of Lyman-$\alpha$ (Ly$\alpha$) emission from galaxies at the onset of cosmic reionization, aiming to understand the conditions necessary for detecting high-redshift sources like JADES-GS-z13-1-LA at $z=13$. By integrating galaxy formation models with detailed intergalactic medium (IGM) reionization simulations, we construct high-redshift galaxy catalogs to model intrinsic Ly$\alpha$ profiles and assess their transmission through the IGM. For a galaxy with $M_{\rm UV}\sim -18.5$ like JADES-GS-z13-1-LA, our fiducial model predicts a Ly$\alpha$ transmission of ${\sim}13$\% and there is a probability of observing Ly$\alpha$ emission with an equivalent width $>40${\AA} of up to 10\%. We also explore how variations in the UV ionizing escape fraction, dependent on host halo mass, impact Ly$\alpha$ detectability. Our findings reveal that reionization morphology significantly influences detection chances -- models where reionization is driven by low-mass galaxies can boost the detection probability to as much as 12\%, while those driven by massive galaxies tend to reduce ionized regions around faint emitters, limiting their detectability. This study underscores the importance of reionization morphology in interpreting high-redshift Ly$\alpha$ observations.
\end{abstract}


%%%%%%%%%%%%%%%%%%%%%%%%%%%%%%%%%%%%%%%%%%%%%%%%%%

%%%%%%%%%%%%%%%%% BODY OF PAPER %%%%%%%%%%%%%%%%%%

\section{Introduction}

Lyman-$\alpha$ (Ly$\alpha$) emission is expected to be heavily suppressed by neutral hydrogen in the intergalactic medium (IGM; \citealt{miraldaescide1998apj...501...15m}). As a result, detecting Ly$\alpha$ emitters (LAEs) in the early Universe is unanticipated and suggests enhanced production of ionizing and/or Lyman-$\alpha$ photons in the first galaxies \citep{Stark2015MNRAS.454.1393S,Witten2024NatAs...8..384W}. The recently reported, stunning Ly$\alpha$ detection from JADES-GS-z13-1-LA \citep{Witstok2024arXiv240816608W} with an equivalent width (EW) of $43^{+15}_{-11}${\AA} from $z=13$, is therefore interpreted as measuring the onset of reionization.

Possible mechanisms to increase the intrinsic Ly$\alpha$ emission include a young, metal-poor stellar population \citep{miraldaescide1998apj...501...15m,Schaerer2003A&A...397..527S} and the presence of an active galactic nucleus (AGN; \citealt{ouchi2018pasj...70s..13o,Sobral2018MNRAS.477.2817S}). 
However, clustered environments can also boost observability when a large ionized region is hollowed out by a cluster of galaxies \citep{qin2022mnras.510.3858q,leonova2022mnras.515.5790l}. Furthermore, outflows within the interstellar medium can shift the intrinsic Ly$\alpha$ line redwards, allowing the emission to penetrate deeper into the damping wing and significantly reducing its cross section before hitting the neutral IGM \citep{Dijkstra2011MNRAS.414.2139D}.

Recent advancements with JWST have substantially expanded the sample of high-redshift LAEs (see e.g., \citealt{Witstok2024A&A...682A..40W}, \citealt{Tang2024arXiv240801507T,jones2024jadesmeasuringreionizationproperties} and references therein). 
With improved statistics to refine the modelling of Ly$\alpha$ profiles \citep{mason2018apj...856....2m},
this work aims to quantify the detectability of Ly$\alpha$ emission from $z=13$. 
By integrating IGM reionization simulations with semi-analytic galaxy-formation models, we use galaxies with realistic UV properties to infer their Ly$\alpha$ line profiles and to self-consistently compute the morphology of neutral hydrogen during the Epoch of Reionization (EoR). Our findings suggest up to a 10\% chance of observing a Ly$\alpha$ equivalent width $>40${\AA} in a galaxy with $M_{\rm UV}\sim -18.5$ like JADES-GS-z13-1-LA  \citep{Witstok2024arXiv240816608W},
increasing to 12\% when varying reionization morphologies are further considered.

After briefly summarizing our model in the next section, we present the results in Section \ref{sec:result} and conclude in Section \ref{sec:conclusion}. In this work, we apply the cosmological parameters from {\it Planck} 2018 ($\Omega_{\mathrm{m}}, \Omega_{\mathrm{b}}, \Omega_{\mathrm{\Lambda}}, h, \sigma_8, n_\mathrm{s} $ = 0.312, 0.0490, 0.688, 0.675, 0.815, 0.968; \citealt{Planck2020A&A...641A...6P}).

\section{Modelling high-redshift Ly$\alpha$ transmission}\label{sec:model}
We model galaxy formation and reionization using the \textit{Meraxes}\footnote{https://github.com/meraxes-devs/meraxes} semi-analytic model \citep{Mutch2016MNRAS.463.3556M,Qin2017a,Balu2023MNRAS.520.3368B}, applied to dark matter halo merger trees that resolve all atomic-cooling halos at $z\leq20$ within a cosmological volume of 210 $h^{-1}$ cMpc per side. The model implements various astrophysical processes to generate a statistically representative sample of high-redshift galaxies, including accretion and cooling of the gaseous reservoir, evolution and feedback of the stellar component and AGN, merger and suppression of the satellite companions, and photoheating during the EoR. The UV and X-ray emissivities are channeled to \textit{21cmFAST} \citep{Mesinger2011MNRAS.411..955M,Murray2020JOSS....5.2582M} to assess the ionization and thermal status of the IGM.

\begin{figure*}
    \centering
    \hspace{-1.5mm} \includegraphics[width=0.4985\textwidth]{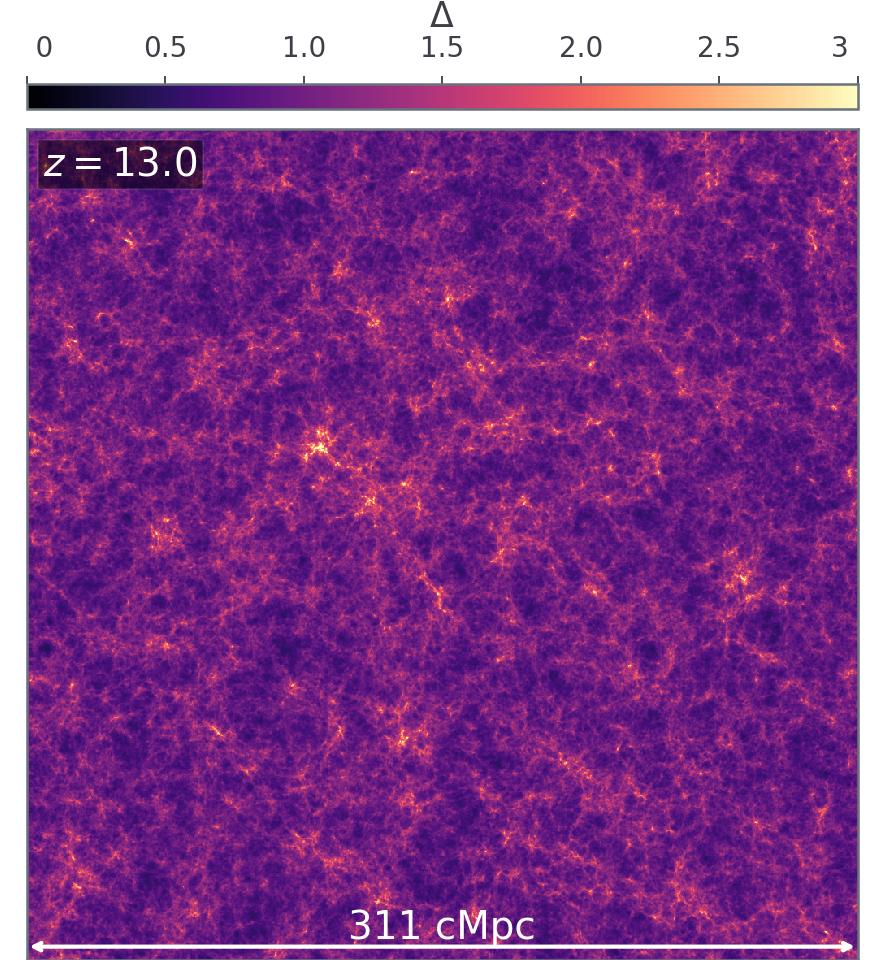}\hspace{-3mm}
    \includegraphics[width=0.493\textwidth]{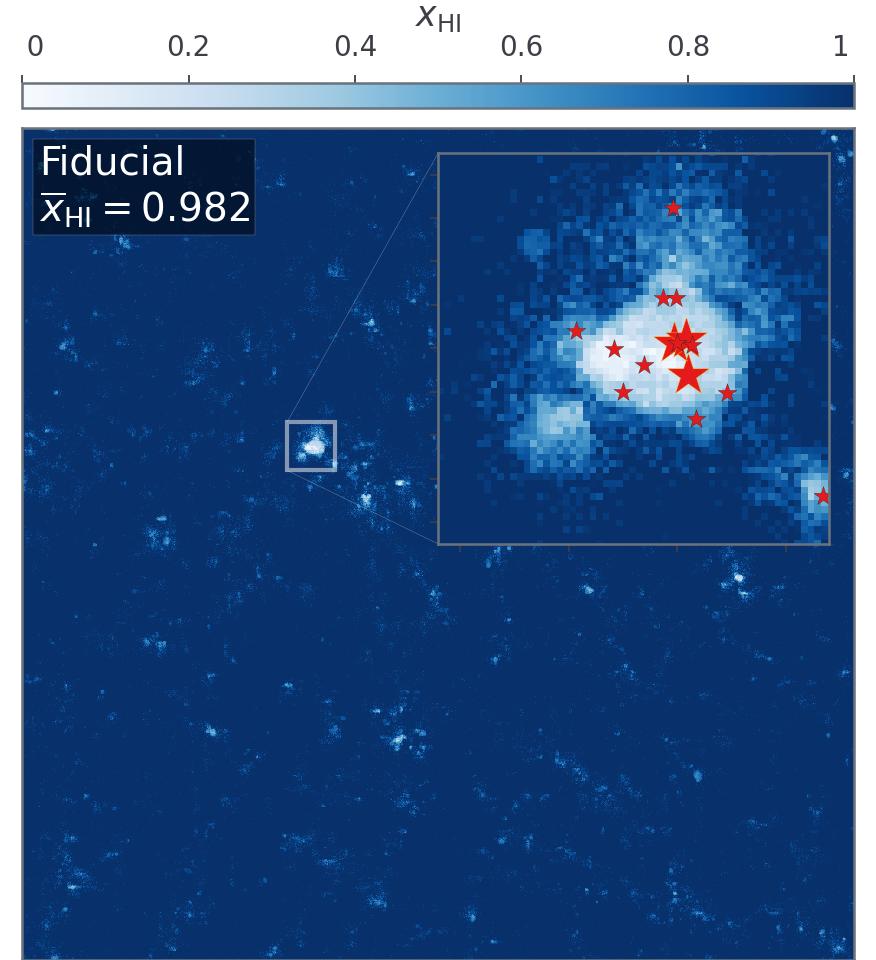}\\
    \includegraphics[width=0.47\textwidth]{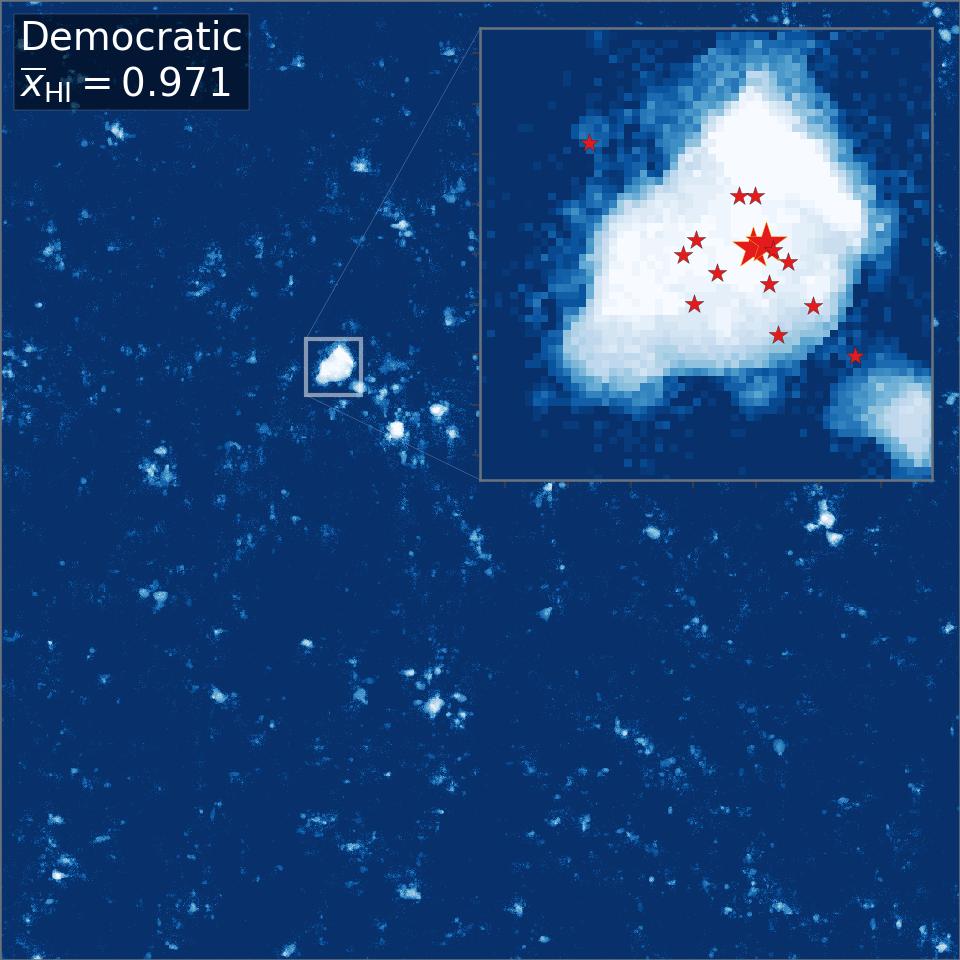} \hspace{0.9mm}
    \includegraphics[width=0.47\textwidth]{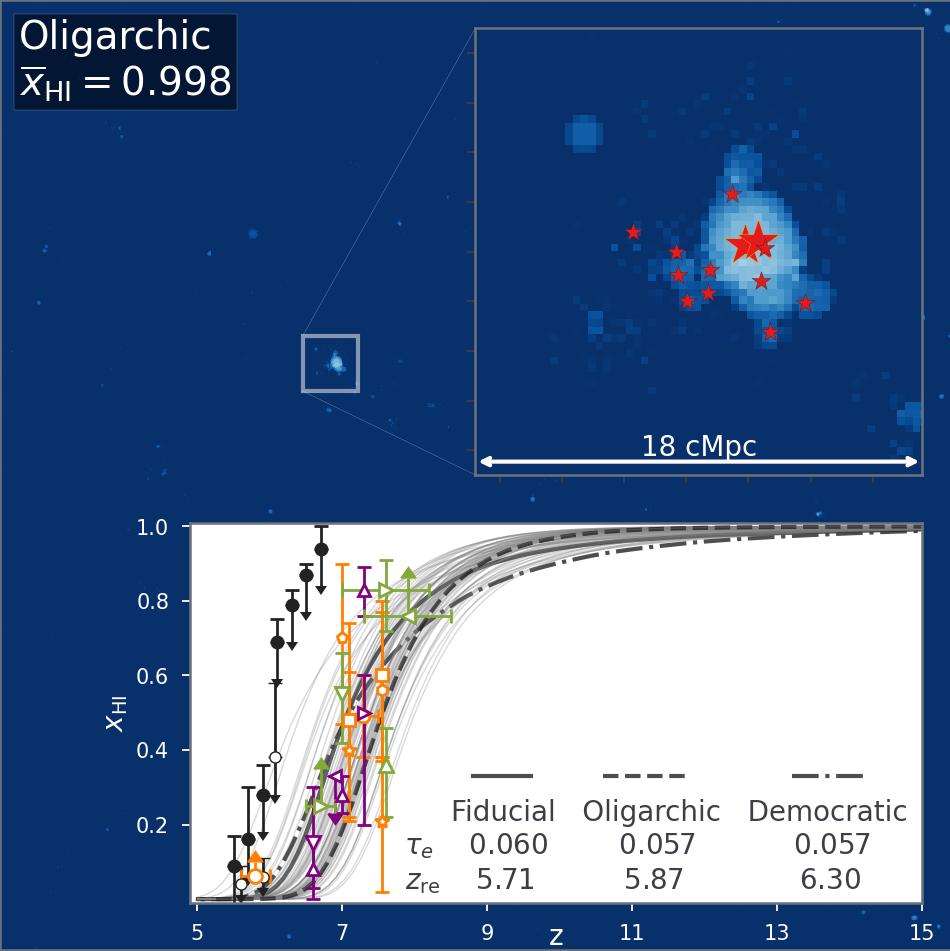}\\
    \caption{\label{fig:HImaps}Slides of overdensity and neutral hydrogen fraction at $z=13$ (projected over a depth of 7cMpc, larger than most ionized bubbles) for three models with different contributions of galaxies to the ionizing budget based on their host halo masses. The upper right zoom-in insets show galaxies brighter than $M_{\rm UV} = -17$~mag as star symbols, with their size representing galaxy brightness. The lower right inset displays the globally averaged reionization histories for these three models (indicating CMB optical depth and reionization timing) along with 70 additional models with varying ionizing escape fractions. These models are broadly consistent with observational results that are based on dark pixel measurements (black; \citealt{McGreer2015MNRAS.447..499M, Jin2023ApJ...942...59J}), damping-wing absorption in quasar spectra (orange; \citealt{Banados2018Natur.553..473B, Greig2017, Greig2019, Greig2022, Davies2018, Wang2020ApJ...896...23W, Zhu2024MNRAS.533L..49Z}), equivalent width measurements (green; \citealt{Mason2019MNRAS.485.3947M, Jung2020ApJ...904..144J, Whitler2020MNRAS.495.3602W, Bolan2022MNRAS.517.3263B}), luminosity functions or clustering of Ly$\alpha$ emitters (purple; \citealt{Inoue2018PASJ...70...55I, Morales2021ApJ...919..120M, ouchi2018pasj...70s..13o, wold2022apj...927...36w}), and CMB polarization (\citealt{Planck2020A&A...641A...6P, qin2020mnras.499..550q}).
    }
\end{figure*}

\begin{table*}
\centering
\caption{Summary of models used in this work. $\mathcal{M}_{10}$ denotes  $\log_{10}\left(M_{\rm vir}/10^{10}{\rm M}_\odot\right)$.
 }\label{tab:models}

\begin{threeparttable}
 \begin{tabular}{l|c|c|cc|l}
 \hline
 \hline
 Model & $\log_{10}v_{\alpha}$& EW& $f_{\rm esc, 10}$ & $\alpha_{\rm esc}$ & description\\%\tnote{*}\\
 \hline
 {\it Fiducial} & $\mathcal{N}\left(0.32 \mathcal{M}_{10} + 2.08,\,0.24^{2}\right)$ &${\rm Exp}\left[100+80\tanh\left(M_{\rm UV}{+}19\right)\right]$ & 0.15 & 0 & better fit of $v_{\alpha}$ and EW to latest LAEs \\
 {\it M18} & $\mathcal{N}\left(0.32 \mathcal{M}_{10} + 1.78,\,0.24^{2}\right)$ &${\rm Exp}\left\{31+12\tanh\left[4\left(M_{\rm UV}{+}20.25\right)\right]\right\}$& \multicolumn{2}{c|}{same as {\it Fiducial}}& $v_{\alpha}$ and EW follow \citet{mason2018apj...856....2m}\\
 {\it LargeScatter} & $\mathcal{N}\left(0.32 \mathcal{M}_{10} + 2.08,\,0.34^{2}\right)$&same as {\it Fiducial} &\multicolumn{2}{c|}{same as {\it Fiducial}} & exploring a larger scatter in $v_{\alpha}$\\
 {\it Democratic} & same as {\it Fiducial} &same as {\it Fiducial} &0.05 & -0.4 & reionization driven by low-mass galaxies\\
 {\it Oligarchic} & same as {\it Fiducial} & same as {\it Fiducial}& 0.45 & 0.9 & reionization driven by high-mass galaxies \\
 ... &same as {\it Fiducial}&same as {\it Fiducial}& ${\leq}0.45$ & $-1.5{\sim}1$ & exploring more reionization morphologies\\
\hline
\end{tabular}
%\begin{tablenotes}
%\item[*] 
%\end{tablenotes}
\end{threeparttable}
\end{table*}

\subsection{Modelling reionization morphology}

Our fiducial model \citep{Qin2023MNRAS.526.1324Q} is calibrated to reproduce well-established galaxy UV luminosity functions measured by HST (e.g. \citealt{Finkelstein2015ApJ...810...71F,Oesch2016ApJ...819..129O,Atek2018MNRAS.479.5184A,Ishigaki2018ApJ...854...73I,Bhatawdekar2019MNRAS.486.3805B,Bouwens2021AJ....162...47B,Bouwens2023MNRAS.523.1036B}) and has shown consistency with recent JWST result up to at least $z\sim13$ (e.g. \citealt{Donnan2023MNRAS.518.6011D,finkelstein2022ApJ...940L..55F,Harikane2023ApJS..265....5H,Naidu2022ApJ...940L..14N,PG2023arXiv230202429P,Willott2024ApJ...966...74W}). Assuming a UV ionizing photon escape fraction ($f_{\rm esc}$) of 15\%, the resulting reionization history aligns with measurements at redshifts between 5 and ${\sim}10$ (e.g. \citealt{McGreer2015MNRAS.447..499M,Banados2018Natur.553..473B,ouchi2018pasj...70s..13o,Wang2020ApJ...896...23W,Jung2020ApJ...904..144J,Whitler2020MNRAS.495.3602W,Morales2021ApJ...919..120M,Greig2022,wold2022apj...927...36w,Jin2023ApJ...942...59J,Zhu2024MNRAS.533L..49Z}). For detailed information on our model and its calibration, we direct interested readers to the referenced publications.

To explore different reionization morphologies at $z=13$, we parameterize $f_{\rm esc}$ as a function of the host halo mass ($M_{\rm vir}$), following the approach of \citet{Park2019MNRAS.484..933P}
\begin{equation}
    f_{\rm esc} = \min\left[1, f_{\rm esc,10} \left(\frac{M_{\rm vir}}{10^{10}{\rm M}_{\odot}}\right)^{\alpha_{\rm esc} }\right],
\end{equation}
where $f_{\rm esc,10}$ and $\alpha_{\rm esc}$ represent the normalization and scaling index, respectively, and are considered as free parameters in this study. The escape fraction is also capped at 100\% in all models. It is important to note that $f_{\rm esc}$ remains highly uncertain, both in observations available at low redshift (e.g., \citealt{Steidel2018ApJ...869..123S,Naidu2018MNRAS.478..791N,Fletcher2019ApJ...878...87F,Izotov2021MNRAS.503.1734I,Pahl2021arXiv210402081P}) and in detailed reionization simulations (e.g., \citealt{ma2020mnras.498.2001m,Kostyuk2023MNRAS.521.3077K,Choustikov2024MNRAS.529.3751C}), with some simulations suggesting increased $f_{\rm esc}$ in less massive galaxies \citep{Paardekooper2015MNRAS.451.2544P,Xu2016ApJ...833...84X,Kostyuk2023MNRAS.521.3077K}. This aligns with expectations that it is easier for supernovae to evacuate low column density channels from shallower gravitational potentials, which allows ionizing photons to escape. Within a Bayesian framework, our systematic exploration of escape fraction variations with halo mass also indicates that current constraints favor larger $f_{\rm esc}$ in lower-mass galaxies (i.e. a negative value of $\alpha_{\rm esc}$; \citealt{mutch2024mnras.527.7924m}). 

We conduct 234 simulations\footnote{Varying $f_{\rm esc,10}$ and $\alpha_{\rm esc}$ alters reionization histories but still has minimal impact on bright observables. Therefore, the models presented in this work remain consistent with the observed luminosity functions, ensuring the drivers of reionization are always statistically representative of the real Universe.} by varying the two ionizing free parameters within the range of $f_{\rm esc,10}\in(0, 0.45]$ and $\alpha_{\rm esc}\in[-1.5,1.0]$. For this work, we only consider 73 of these simulations, which meet the criteria of having a cosmic microwave background (CMB) optical depth consistent with {\it Planck} measurements (i.e. $0.05\leq\tau_e\leq0.065$; \citealt{Planck2020A&A...641A...6P,qin2020mnras.499..550q}) and reaching a globally averaged neutral fraction of 1\% after $z_{\rm re}=6.8$ (broadly consistent with various observations, e.g., \citealt{Banados2018Natur.553..473B,Whitler2020MNRAS.495.3602W,Morales2021ApJ...919..120M,Jin2023ApJ...942...59J}; see their reionization histories in the lower-right panel of Fig. \ref{fig:HImaps}). Although these simulations exhibit similar reionization histories, they show significant variations in the spatial distribution of neutral hydrogen. We use these simulations to investigate how reionization morphology affects the detectability of Ly$\alpha$ emission in the highly neutral universe at $z=13$. 

Fig. \ref{fig:HImaps} highlights three example models: (i) {\it Fiducial} with $f_{\rm esc}=0.15$; (ii) {\it Democratic} with $f_{\rm esc,10}=0.05$ and $\alpha_{\rm esc}=-0.4$; and (iii) {\it Oligarchic} with $f_{\rm esc,10}=0.45$ and $\alpha_{\rm esc}=0.9$, are summarized in Table \ref{tab:models}. There are notable differences in reionization morphology between these models. Compared to the {\it Fiducial} model, {\it Democratic} allows less-massive galaxies to emit more ionizing photons into the IGM and, due to the higher abundance of low-mass galaxies at high redshifts, {\it Democratic} features a greater number of small-sized ionized bubbles at $z=13$. Additionally, overdense regions (see the zoom-in insets) can expand their surrounding HII regions with the aid of clustered low-mass galaxies. This model aligns most closely with the escape fraction constraints from \citet{mutch2024mnras.527.7924m}.
In contrast, the {\it Oligarchic} model, which relies on galaxies with well-established stellar mass to ionize the IGM, shows delayed reionization due to the scarcity of these massive contributors in the early stages of the EoR. Furthermore, as illustrated by the (dis)appearance of galaxies in Fig. \ref{fig:HImaps}, the brightness of the first galaxies is sensitive to the ionization history and can be significantly influenced by local photoheating feedback. 

The globally averaged neutral fraction also evolves differently between these models: {\it Democratic} shows a more extended and slowly evolving neutral fraction, while {\it Oligarchic} exhibits a narrower and more rapid evolution compared to the {\it Fiducial} model. Despite these variations, the globally averaged neutral fraction $\overline{x}_{\rm HI}$ at $z=13$ remains similar across models (all above 97\%), with a CMB optical depth around $\tau_e = 0.06$ and reionization concluding around $z_{\rm re} = 6$.

\subsection{Modelling the Ly$\alpha$ line profile}

We model the Ly$\alpha$ profile ($F_\alpha$) before damping-wing absorption as a truncated normal distribution with respect to velocity ($v$): $F_\alpha \sim \mathcal{N}\left(v_{\alpha}, \frac{v_{\alpha}^2}{\ln 2^8}\right)$ for $v > v_{\rm circ}$, and 0 otherwise (see Fig. 1 in \citealt{mason2018apj...856....2m} for examples). Here, the full width at half maximum of the profile is assumed to be $v_{\alpha}$, and emission blueward of the circular velocity of the host halo ($v_{\rm circ}$) is considered completely absorbed by residual neutral hydrogen within ionized bubbles. To model the Ly$\alpha$ damping-wing transmission, we follow \citet{mason2018apj...856....2m} and sample the intrinsic Ly$\alpha$ line offset ($v_{\alpha}$) and equivalent width (EW) with slight modifications to their originally proposed distributions.

\begin{figure}
%    \centering
    \includegraphics[width=\columnwidth]{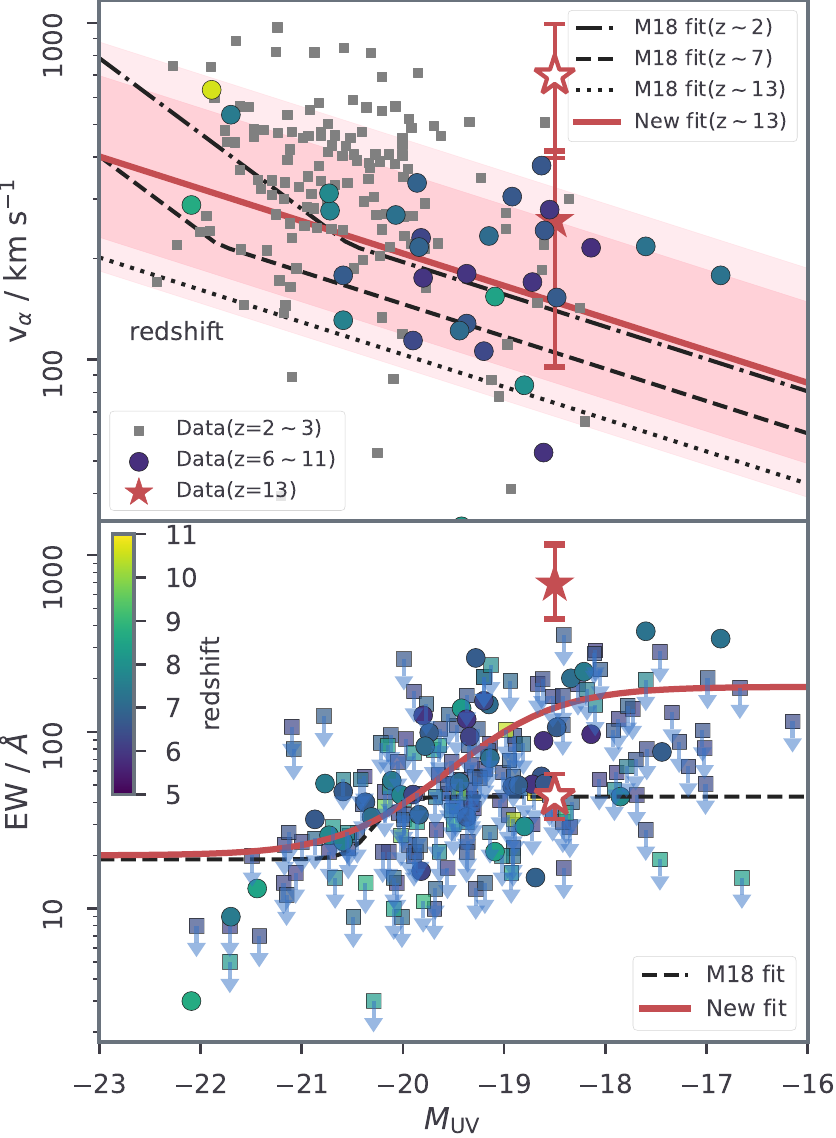}
    \caption{\label{fig:obs} Ly$\alpha$ velocity offset ($v_{\alpha}$) and equivalent width (EW) as a function of UV magnitude. Observational data are categorized by redshift ranges: $z=2\sim3$ (from \citealt{mason2018apj...856....2m} and references therein), $z=6\sim11$ (from \citealt{Witstok2024A&A...682A..40W}, \citealt{Tang2024arXiv240801507T} and references therein), and $z=13$ (\citealt{Witstok2024arXiv240816608W}). At $z=13$ the open and filled stars respectively represent the observed and predicted intrinsic values following spectral fitting including Ly$\alpha$ and IGM and DLA transmission (\citealt{Witstok2024arXiv240816608W}). In the top panel, the curves represent the mean velocity offset $v_{\alpha}$ of the original log-normal distribution fitted by \citet{mason2018apj...856....2m} for LAEs at $z \sim 2$, 7, and 13, along with our newly proposed fit. The dark and light shaded regions indicate the variance reported by \citet{mason2018apj...856....2m} and the one further explored in this work. The scale parameter of the fitted exponential distributions for EW is shown in the bottom panel.}
\end{figure}

\citet{mason2018apj...856....2m} created an empirical model for the $v_{\alpha}$--$M_{\rm UV}$ relation by first converting UV magnitude to halo mass using the following relation \citep{mason2015}
\begin{equation}
\log_{10}\left(\frac{M_{\rm vir}}{10^{10}{\rm M}_\odot}\right) = \gamma \left(M_{\rm UV} - M_{\rm UV,c}\right) +1.75,
\end{equation}
where $M_{\rm UV,c}=-20-0.26z$ and $\gamma=-0.7(-0.3)$ for galaxies brighter(fainter) than $M_{\rm UV,c}$. \citet{mason2018apj...856....2m} then used measurements of LAEs, primarily from post-reionization ($z\sim2$--3, based on {\it Keck}; e.g., \citealt{Erb2014ApJ...795...33E}), to fit Ly$\alpha$ line offset using a log-normal distribution with a variable mean that depends on the host halo mass yielding
 \begin{equation}
 \log_{10} v_{\alpha} \sim \mathcal{N}\left[m \log_{10}\left(\frac{M_{\rm vir}}{10^{10} {\rm M}_\odot}\right)    + c, \sigma^{2}\right],
\end{equation}
where parameters $(m, c, \sigma)$ = $(0.32, 1.78, 0.24)$. The top panel of Fig. \ref{fig:obs} shows this low-redshift sample in grey and plots their fitted mean $v_{\alpha}$ against UV magnitude, indicated the long dash-dotted curve. 

The top panel of Fig. \ref{fig:obs} also presents the resulting predicted mean $v_{\alpha}-M_{\rm UV}$ relation at $z \sim 7$ and 13,
along with the latest LAE measurements from JWST (see \citealt{Witstok2024A&A...682A..40W,Tang2024arXiv240801507T} and references therein), 
including the new $z = 13$ detection by \citet{Witstok2024arXiv240816608W}. In their analysis \citet{mason2018apj...856....2m} noted negligible differences in the $v_{\alpha}-M_{\rm vir}$ distribution between the low-redshift sample and the observations at $z>6$ that were available at the time (e.g., \citealt{Stark2017MNRAS.464..469S}), and therefore modelled a mean $v_{\alpha}$–$M_{\rm vir}$ relation that does not evolve with redshift. 
%\footnote{Since the Ly$\alpha$ velocity offset is mainly determined by small-scale ISM properties such as HI column density and temperature (thermal brocading), outflows (bulk motion) and dust content (scattering), its correlation with global galaxy properties is uncertain. However, structure formation driven by gravity (e.g., the stellar-to-halo-mass ratio being primarily a function of halo mass and independent of redshift at $z \geq 6$; \citealt{stefanon2021apj...922...29s}) motivates a redshift-independent $v_{\alpha}$–$M_{\rm vir}$ relation.}. 
However, newer data indicates that the redshift-independent $v_{\alpha}$–$M_{\rm vir}$ relation requires revision to better fit the current LAE sample at higher redshifts. Indeed, if the relation between $v_{\alpha}$ and halo mass were driven by the depth of the gravitational potential well we would expect an offset at fixed halo mass that increased towards higher redshift. Therefore, we propose increasing the intercept by 0.3 (to a value of $c=2.08$), and show the new fit for $z \sim 13$ in the top panel of Fig. \ref{fig:obs} for comparison. In addition to the original fit proposed by \citet{mason2018apj...856....2m} and this new fit with a larger $c$, we also explore a larger variance of $\sigma = 0.34$, which corresponds to the standard deviation of the high-redshift sample shown in the panel. Although this choice is somewhat arbitrary, it aims to account for the increased scatter observed in the new high-redshift data. We refer to these three models as {\it M18}, {\it Fiducial}, and {\it LargeScatter} and summarize them in Table \ref{tab:models}.

As shown in the bottom panel of Fig. \ref{fig:obs}, much more observational data for the EW of high-redshift LAEs has become available since the study by \citet{deboer2017pasp..129d5001d}, whose sample was used by \citet{mason2018apj...856....2m} to fit an exponential distribution (see also \citealt{Oyarzun2017ApJ...843..133O}) for $z \sim 6$ LAEs
\begin{equation}
    {\rm EW}\sim {\rm Exp}\left\{{\rm EW}_{\rm c}+j\tanh\left[k\left(M_{\rm UV}-M_{\rm UV,c}\right)\right]\right\},
\end{equation}
with $({\rm EW}_{\rm c}, j, k, M_{\rm UV,c}) = (31, 12, 4, -20.25)$. This fit was motivated by splitting the LAE sample into brighter than $M_{\rm UV} = -21$~mag, fainter than $-20$ mag, and in-between, as well as by providing a smooth transition between $M_{\rm UV} = -21$ and $-20$ mag. Since this work focuses on much fainter luminosities, we adopt a new fit with $({\rm EW}_{\rm c}, j, k, M_{\rm UV,c}) = (100, 80, 1, -19)$ for the {\it Fiducial} and {\it LargeScatter} models. As illustrated in the panel, our new fit agrees better with LAEs of higher redshifts and fainter luminosities while remaining comparable to \citet{mason2018apj...856....2m} at the bright end.

Notably, recent measurements of the Ly$\alpha$ forest indicate that reionization may have occurred later than $z = 6$, with significant fluctuations in the photoionizing background during these early times \citep{nasir2020mnras.494.3080n,qin2021mnras.506.2390q,bosman2022mnras.514...55b,Gaikwad2023MNRAS.525.4093G,Davies2024ApJ...965..134D}.
Therefore, these newly observed LAEs from JWST might be attenuated, as the damping-wing transmissions ($\mathcal{T}$) at their redshifts are likely lower than 100\%. Consequently EWs sampled from the aforementioned fits might still be lower than the intrinsic values. To account for this, for each simulation\footnote{In practice, for each of the 500 realizations of a simulation, we additionally sample a set of $v_{\alpha}$ and IGM skews for all modeled galaxies in the $z = 6$ snapshot. We conservatively employ the original fit from \citet{mason2018apj...856....2m} for all models listed in Table \ref{tab:models} and compute the transmission, $\mathcal{T}(z = 6)$, following Sec. \ref{subsec:transmission}. As we find the distribution of $\mathcal{T}(z = 6)$ is highly similar for galaxies within the three luminosity bins studied here, we build a probability distribution function of $\mathcal{T}(z = 6) \sim \mathcal{L}$ using all galaxies within $-20 < M_{\rm UV} \le -17$. When de-absorbing, we randomly draw from $\mathcal{L}$ and divide the modeled EWs at higher redshifts by $\mathcal{T}(z = 6)$.} we use the $z = 6$ snapshot to de-absorb the sampled EWs at higher redshifts. Dropping the assumption of $\mathcal{T}(z \sim 6) = 1$ means that models with varying EoR endings but sharing the same transmission at higher redshifts will predict different LAE detectabilities, thus breaking degeneracy in reionization modeling.

\subsection{Modelling the Ly$\alpha$ observable}\label{subsec:transmission}

\begin{figure*}
    \centering
    \includegraphics[width=0.8\textwidth]{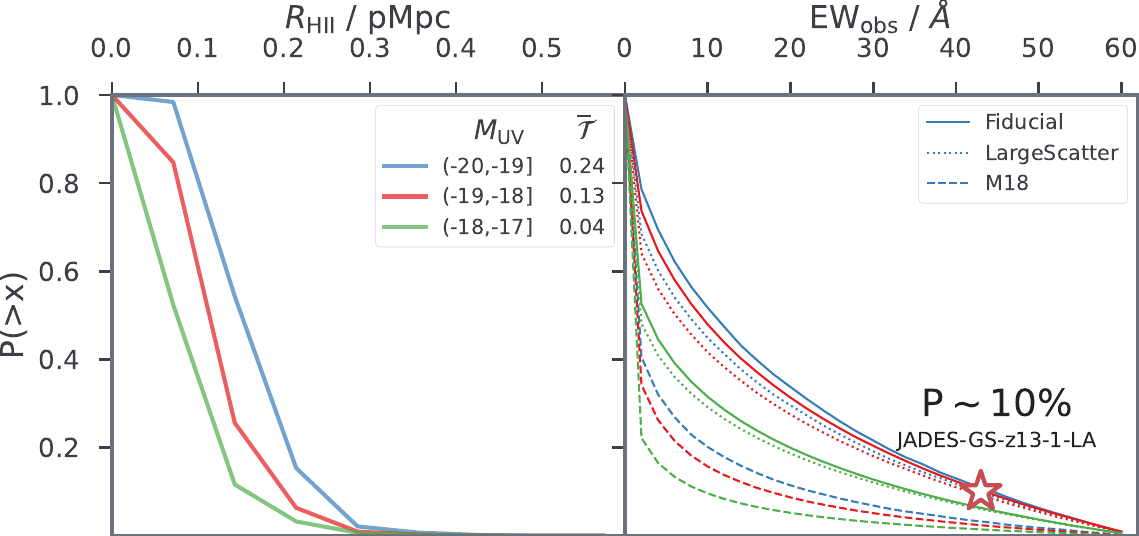}\\\vspace*{-0.5mm}
    \includegraphics[width=0.8\textwidth]{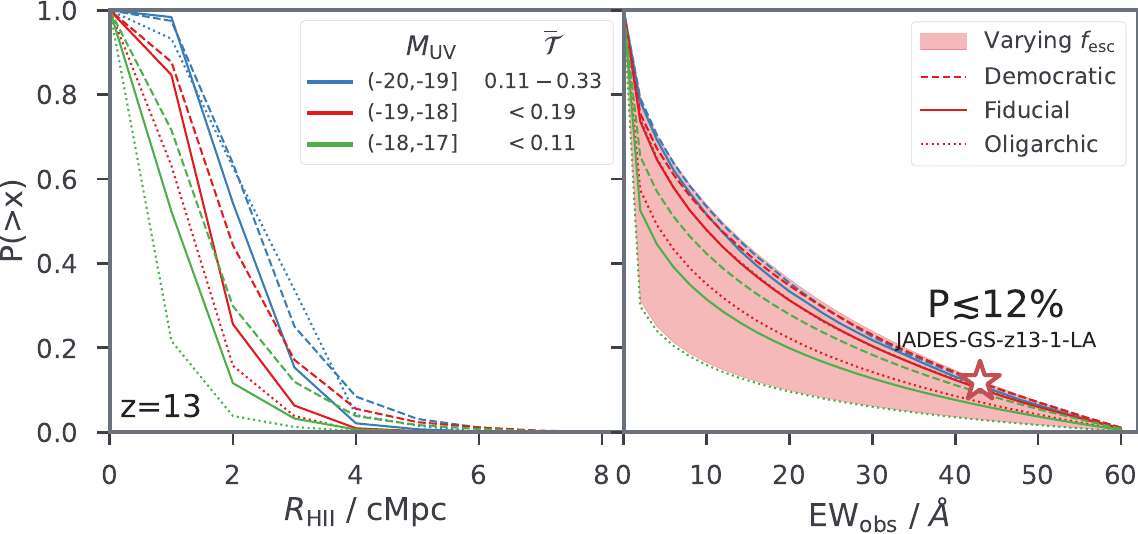}%\vspace*{-2mm}
    
    \caption{Probability that a galaxy in three UV magnitude bins (different colors) is located in an ionized region larger than $R_{\rm HII}$ (left panels) or has an observed Ly$\alpha$ emission greater than EW$_{\rm obs}$ (right panels). The top panels compare different Ly$\alpha$ sampling models ({\it M18}, {\it Fiducial} and {\it LargeScatter}) while the bottom panels assess the impact of reionization morphology ({\it Fiducial}, {\it Democratic} and {\it Oligarchic}). In the lower-right panel, the shaded region encapsulate the EW$_{\rm obs}$ distribution across all 73 models used in this work for $-19<M_{\rm UV}\leq-18$. 
    Tabulated in the upper left panel are the median Ly$\alpha$ transmissions, $\overline{\mathcal{T}}$, for all $z=13$ galaxies in the {\it Fiducial} model within each of the $M_{\rm UV}$ bins. Correspondingly, the range of $\overline{\mathcal{T}}$ for all 73 varying-$f_{\rm esc}$ models is listed in the lower left panel. 
    \label{fig:cdf}}
\end{figure*}

For each galaxy, we perform 500 random realizations with each drawing (i) an intrinsic velocity offset from the log-normal distribution; (ii) an EW based on the exponential distribution; and (iii) a line of sight in the IGM to obtain skews of the local density ($\Delta$) and neutral fraction ($x_{\rm HI}$), which also allows us to measure size of the ionized region ($R_{\rm HII}$). We then compute the damping-wing optical depth outside the HII region using the following integral over distance ($l$):
\begin{equation}
    \tau_{\rm HI} = \int {\rm d}l \Delta x_{\rm HI} \overline{n}_{\rm H} \sigma_\alpha.
\end{equation}
Here, $\overline{n}_{\rm H}$ represents the cosmic mean hydrogen density at the time, and the Ly$\alpha$ cross-section is evaluated as \citep{miraldaescide1998apj...501...15m}:
\begin{equation}\label{eq:sigma_lya}
    \sigma_{\alpha}\left(x{\equiv}\frac{\lambda_\alpha}{\lambda+\lambda_\alpha{\rm v}_\alpha/c}\right) = \frac{3\lambda_{\alpha}^2}{8\pi} \frac{x^4}
	{1/4x^6 + (1-x)^2({2\pi c}/{\lambda_\alpha\Lambda})^2},
\end{equation}
where $c$, $\Lambda$, $\lambda_\alpha$ and $\lambda$ are the speed of light, decay constant for Ly$\alpha$ resonance, Ly$\alpha$ line centre and its red side, respectively. Finally, the Ly$\alpha$ damping-wing transmission ($\mathcal{T}$) is given by 
\begin{equation}
    \mathcal{T} = \int F_\alpha(v) e^{-\tau_{\rm HI}(v)}
\end{equation}
and the observed EW is simply ${\rm EW}_{\rm obs}(z)={\rm EW} \times \mathcal{T}(z) / \mathcal{T}(z=6)$.

\section{Detecting LAEs at $z=13$}\label{sec:result}

Our results are summarized in Fig. \ref{fig:cdf}, which shows the probability that a galaxy with $M_{\rm UV}\sim$-19.5, -18.5, or -17.5 is located in an ionized region larger than $R_{\rm HII}$ (left panels) or has an observed Ly$\alpha$ emission greater than EW$_{\rm obs}$ (right panels). In the top panels, we compare the models {\it M18}, {\it Fiducial} and {\it LargeScatter}, which share the same UV ionizing fraction but differ in their Ly$\alpha$ modelling. As reionization morphology does not change between these models, they share the same bubble size distribution as shown in the upper left panel. We observe that fainter galaxies generally reside in smaller ionized regions but can sometimes be found in larger HII bubbles (up to $\sim 8$cMpc) when clustered around brighter neighbours. This is illustrated in the zoom-in insets of Fig. \ref{fig:HImaps} and also studied by \citet{qin2022mnras.510.3858q} in more detail.

On the other hand, the Ly$\alpha$ detectability predictions differ among {\it M18}, {\it Fiducial} and {\it LargeScatter}. 
For galaxies with UV luminosities similar to JADES-GS-z13-1-LA (${\sim}$-18.5 mag), our {\it Fiducial} model predicts a median Ly$\alpha$ transmission of around 13\%. When further restricting the modelled galaxies to those with the observed EWs of at least $\sim$40{\AA} -- matching that of the detected LAE\footnote{The measured line offset has greater uncertainties and therefore are not considered when selecting analogues of JADES-GS-z13-1-LA.} -- we find a detectability rate of up to 10\%. 
The detectability of LAEs with a given EW decreases at lower luminosities (i.e. $-18<M_{\rm UV}\leq-17$), as most reside in smaller ionized regions and suffer from a reduced median transmission of only 4\% due to damping-wing absorption. On the other hand, despite having a higher median transmission of 24\%, galaxies brighter than $M_{\rm UV}\sim-19$ mag still show the same maximum detectability of 10\% because their sampled EWs in {\it Fiducial} are preferentially lower (see the bottom panel of Fig. \ref{fig:obs}). Comparing {\it Fiducial} and {\it LargeScatter}, we see that increasing the variance when sampling the velocity offset results in a minor decrease in detectability. This small difference illustrates that our results are not qualitatively dependent on the assumed scatter. Our revised fit increases the probability of detection relative to the fit of {\it M18} \citep{mason2018apj...856....2m}. However, should the $v_{\alpha}$ distribution indeed follow {\it M18}, there is still approximately a 5\% chance of detecting LAEs brighter than EW$\sim30${\AA} at these high redshifts. 

We further probe the impact of reionization morphology on Ly$\alpha$ detectability in the bottom panels of Fig. \ref{fig:cdf}. The three example models shown in Fig. \ref{fig:HImaps} -- {\it Fiducial}, {\it Democratic} and {\it Oligarchic} -- are highlighted. Our findings indicate that, at this early universe, reionization driven by high(low)-mass galaxies leads to smaller(larger) ionized regions around faint galaxies ($M_{\rm UV}\gtrsim-19$). However, as bright galaxies tend to reside in overdense regions, boosting the ionizing fraction of galaxies either smaller or larger than $M_{\rm vir}=10^{10}{\rm M}_\odot$ can advance their local ionization. As a result, uncertainties in ionized bubble size and observed Ly$\alpha$ equivalent width distributions become significantly larger when reionization is driven by galaxies with varying characteristic masses. Assuming all morphologies share an equal likelihood when compared to existing reionization constraints, we illustrate the range of possible detectabilities using the shaded region in the lower-right panel of Fig. \ref{fig:cdf}, encapsulating the distributions across all 73 models with varying $f_{\rm esc}$, for $-19<M_{\rm UV}\leq-18$. While most reionization morphologies reduce the likelihood of detecting JADES-GS-z13-1-LA (as low as 3\%), models with characteristics similar to {\it Democratic} and having low-mass galaxies to drive reionization, can increase the detection probability to around 12\%. 
\section{Conclusions}\label{sec:conclusion}

In this study, we explored the detectability of Ly$\alpha$ emission from galaxies during the onset of reionization. By integrating galaxy formation models with detailed IGM reionization simulations, we constructed high-redshift galaxy catalogues that are statistically representative of the real Universe. These catalogues enable us to empirically model Ly$\alpha$ profiles for galaxies at $z=13$, considering key intrinsic properties like velocity offset and equivalent width (EW).

Our results indicate that while Ly$\alpha$ emission is attenuated by neutral hydrogen in the IGM, specific conditions such as large ionized regions around clustered galaxies and highly redshifted intrinsic emission can enable the detection of LAEs, even at these early cosmic epochs. For a galaxy as bright as JADES-GS-z13-1-LA \citep{Witstok2024arXiv240816608W}, with a UV magnitude of approximately -18.5, we estimate a Ly$\alpha$ transmission of $\sim$13\%. Additionally, under fiducial conditions there is a $\sim$10\% probability of detecting its Ly$\alpha$ emission with an EW of $\sim40${\AA}, matching that of JADES-GS-z13-1-LA.

Reionization morphology plays a crucial role in shaping Ly$\alpha$ visibility. Our analysis shows that when reionization is driven by low-mass galaxies (e.g., the {\it Democratic} model), the probability of detecting JADES-GS-z13-1-LA-like LAEs can increase to as high as 12\%. In contrast, reionization driven by massive galaxies tends to shrink ionized regions around fainter galaxies, further limiting their detectability. We find that fainter galaxies, which generally reside in smaller ionized regions, experience stronger damping-wing absorption and lower Ly$\alpha$ transmission. Conversely, brighter galaxies, despite having larger ionized regions and higher transmission, remain constrained by their lower intrinsic EWs, limiting their overall detectability.

Our modeling indicates that around 1 in 10 galaxies is expected to be observed in  Ly$\alpha$, even at these very high redshifts. At the time of writing there are ${\sim}15$ Lyman-break candidates with $z\in[12,14]$ and $M_{\rm UV}\in[-19,-18]$ (see Fig. 2 of \citealt{Carniani2024arXiv240518485C} and references therein). Of these candididates, JADES-GS-z13-1-LA \citep{Witstok2024arXiv240816608W} is the only galaxy observed as a LAE, with another two spectroscopically confirmed galaxies showing no sign of Ly$\alpha$ \citep{Curtis-Lake2023NatAs...7..622C}.  This work highlights the complexities in detecting Ly$\alpha$ emission from the early Universe and underscores the importance of considering both galaxy properties and reionization morphology when interpreting these high-redshift observations. As the sample of LAEs identified by JWST continues to grow, this study offers valuable insights into the conditions required for early Ly$\alpha$ emission detection and their implications for the onset of cosmic reionization.

\section*{Acknowledgements}
This work was performed on OzSTAR and Gadi in Australia. YQ acknowledges HPC resources from the ASTAC Large Programs and support from the ARC Discovery Early Career Researcher Award (DECRA) through fellowship \#DE240101129.

%%%%%%%%%%%%%%%%%%%%%%%%%%%%%%%%%%%%%%%%%%%%%%%%%%
\section*{Data Availability}
The data underlying this article will be shared on reasonable request to the corresponding author. 
 


\bibliographystyle{mnras}
\bibliography{reference} % if your bibtex file is called example.bib

\begin{thebibliography}{}
\makeatletter
\relax
\def\mn@urlcharsother{\let\do\@makeother \do\$\do\&\do\#\do\^\do\_\do\%\do\~}
\def\mn@doi{\begingroup\mn@urlcharsother \@ifnextchar [ {\mn@doi@}
  {\mn@doi@[]}}
\def\mn@doi@[#1]#2{\def\@tempa{#1}\ifx\@tempa\@empty \href
  {http://dx.doi.org/#2} {doi:#2}\else \href {http://dx.doi.org/#2} {#1}\fi
  \endgroup}
\def\mn@eprint#1#2{\mn@eprint@#1:#2::\@nil}
\def\mn@eprint@arXiv#1{\href {http://arxiv.org/abs/#1} {{\tt arXiv:#1}}}
\def\mn@eprint@dblp#1{\href {http://dblp.uni-trier.de/rec/bibtex/#1.xml}
  {dblp:#1}}
\def\mn@eprint@#1:#2:#3:#4\@nil{\def\@tempa {#1}\def\@tempb {#2}\def\@tempc
  {#3}\ifx \@tempc \@empty \let \@tempc \@tempb \let \@tempb \@tempa \fi \ifx
  \@tempb \@empty \def\@tempb {arXiv}\fi \@ifundefined
  {mn@eprint@\@tempb}{\@tempb:\@tempc}{\expandafter \expandafter \csname
  mn@eprint@\@tempb\endcsname \expandafter{\@tempc}}}

\bibitem[\protect\citeauthoryear{{Atek}, {Richard}, {Kneib}  \&
  {Schaerer}}{{Atek} et~al.}{2018}]{Atek2018MNRAS.479.5184A}
{Atek} H.,  {Richard} J.,  {Kneib} J.-P.,   {Schaerer} D.,  2018, \mn@doi
  [\mnras] {10.1093/mnras/sty1820}, \href
  {https://ui.adsabs.harvard.edu/abs/2018MNRAS.479.5184A} {479, 5184}

\bibitem[\protect\citeauthoryear{{Ba{\~n}ados} et~al.,}{{Ba{\~n}ados}
  et~al.}{2018}]{Banados2018Natur.553..473B}
{Ba{\~n}ados} E.,  et~al., 2018, \mn@doi [\nat] {10.1038/nature25180}, \href
  {https://ui.adsabs.harvard.edu/abs/2018Natur.553..473B} {553, 473}

\bibitem[\protect\citeauthoryear{{Balu}, {Greig}, {Qiu}, {Power}, {Qin},
  {Mutch}  \& {Wyithe}}{{Balu} et~al.}{2023}]{Balu2023MNRAS.520.3368B}
{Balu} S.,  {Greig} B.,  {Qiu} Y.,  {Power} C.,  {Qin} Y.,  {Mutch} S.,
  {Wyithe} J. S.~B.,  2023, \mn@doi [\mnras] {10.1093/mnras/stad281}, \href
  {https://ui.adsabs.harvard.edu/abs/2023MNRAS.520.3368B} {520, 3368}

\bibitem[\protect\citeauthoryear{{Bhatawdekar}, {Conselice},
  {Margalef-Bentabol}  \& {Duncan}}{{Bhatawdekar}
  et~al.}{2019}]{Bhatawdekar2019MNRAS.486.3805B}
{Bhatawdekar} R.,  {Conselice} C.~J.,  {Margalef-Bentabol} B.,   {Duncan} K.,
  2019, \mn@doi [\mnras] {10.1093/mnras/stz866}, \href
  {https://ui.adsabs.harvard.edu/abs/2019MNRAS.486.3805B} {486, 3805}

\bibitem[\protect\citeauthoryear{{Bolan} et~al.,}{{Bolan}
  et~al.}{2022}]{Bolan2022MNRAS.517.3263B}
{Bolan} P.,  et~al., 2022, \mn@doi [\mnras] {10.1093/mnras/stac1963}, \href
  {https://ui.adsabs.harvard.edu/abs/2022MNRAS.517.3263B} {517, 3263}

\bibitem[\protect\citeauthoryear{{Bosman} et~al.,}{{Bosman}
  et~al.}{2022}]{bosman2022mnras.514...55b}
{Bosman} S. E.~I.,  et~al., 2022, \mn@doi [\mnras] {10.1093/mnras/stac1046},
  \href {https://ui.adsabs.harvard.edu/abs/2022MNRAS.514...55B} {514, 55}

\bibitem[\protect\citeauthoryear{{Bouwens} et~al.,}{{Bouwens}
  et~al.}{2021}]{Bouwens2021AJ....162...47B}
{Bouwens} R.~J.,  et~al., 2021, \mn@doi [\aj] {10.3847/1538-3881/abf83e}, \href
  {https://ui.adsabs.harvard.edu/abs/2021AJ....162...47B} {162, 47}

\bibitem[\protect\citeauthoryear{{Bouwens} et~al.,}{{Bouwens}
  et~al.}{2023}]{Bouwens2023MNRAS.523.1036B}
{Bouwens} R.~J.,  et~al., 2023, \mn@doi [\mnras] {10.1093/mnras/stad1145},
  \href {https://ui.adsabs.harvard.edu/abs/2023MNRAS.523.1036B} {523, 1036}

\bibitem[\protect\citeauthoryear{{Carniani} et~al.,}{{Carniani}
  et~al.}{2024}]{Carniani2024arXiv240518485C}
{Carniani} S.,  et~al., 2024, \mn@doi [arXiv e-prints]
  {10.48550/arXiv.2405.18485}, \href
  {https://ui.adsabs.harvard.edu/abs/2024arXiv240518485C} {p. arXiv:2405.18485}

\bibitem[\protect\citeauthoryear{{Choustikov} et~al.,}{{Choustikov}
  et~al.}{2024}]{Choustikov2024MNRAS.529.3751C}
{Choustikov} N.,  et~al., 2024, \mn@doi [\mnras] {10.1093/mnras/stae776}, \href
  {https://ui.adsabs.harvard.edu/abs/2024MNRAS.529.3751C} {529, 3751}

\bibitem[\protect\citeauthoryear{{Curtis-Lake} et~al.,}{{Curtis-Lake}
  et~al.}{2023}]{Curtis-Lake2023NatAs...7..622C}
{Curtis-Lake} E.,  et~al., 2023, \mn@doi [Nature Astronomy]
  {10.1038/s41550-023-01918-w}, \href
  {https://ui.adsabs.harvard.edu/abs/2023NatAs...7..622C} {7, 622}

\bibitem[\protect\citeauthoryear{Davies et~al.,}{Davies
  et~al.}{2018}]{Davies2018}
Davies F.~B.,  et~al., 2018, \mn@doi [The Astrophysical Journal]
  {10.3847/1538-4357/aad6dc}, 864, 142

\bibitem[\protect\citeauthoryear{{Davies} et~al.,}{{Davies}
  et~al.}{2024}]{Davies2024ApJ...965..134D}
{Davies} F.~B.,  et~al., 2024, \mn@doi [\apj] {10.3847/1538-4357/ad1d5d}, \href
  {https://ui.adsabs.harvard.edu/abs/2024ApJ...965..134D} {965, 134}

\bibitem[\protect\citeauthoryear{{DeBoer} et~al.,}{{DeBoer}
  et~al.}{2017}]{deboer2017pasp..129d5001d}
{DeBoer} D.~R.,  et~al., 2017, \mn@doi [\pasp]
  {10.1088/1538-3873/129/974/045001}, \href
  {https://ui.adsabs.harvard.edu/abs/2017PASP..129d5001D} {129, 045001}

\bibitem[\protect\citeauthoryear{{Dijkstra}, {Mesinger}  \&
  {Wyithe}}{{Dijkstra} et~al.}{2011}]{Dijkstra2011MNRAS.414.2139D}
{Dijkstra} M.,  {Mesinger} A.,   {Wyithe} J. S.~B.,  2011, \mn@doi [\mnras]
  {10.1111/j.1365-2966.2011.18530.x}, \href
  {https://ui.adsabs.harvard.edu/abs/2011MNRAS.414.2139D} {414, 2139}

\bibitem[\protect\citeauthoryear{{Donnan} et~al.,}{{Donnan}
  et~al.}{2023}]{Donnan2023MNRAS.518.6011D}
{Donnan} C.~T.,  et~al., 2023, \mn@doi [\mnras] {10.1093/mnras/stac3472}, \href
  {https://ui.adsabs.harvard.edu/abs/2023MNRAS.518.6011D} {518, 6011}

\bibitem[\protect\citeauthoryear{{Erb} et~al.,}{{Erb}
  et~al.}{2014}]{Erb2014ApJ...795...33E}
{Erb} D.~K.,  et~al., 2014, \mn@doi [\apj] {10.1088/0004-637X/795/1/33}, \href
  {https://ui.adsabs.harvard.edu/abs/2014ApJ...795...33E} {795, 33}

\bibitem[\protect\citeauthoryear{{Finkelstein} et~al.,}{{Finkelstein}
  et~al.}{2015}]{Finkelstein2015ApJ...810...71F}
{Finkelstein} S.~L.,  et~al., 2015, \mn@doi [\apj]
  {10.1088/0004-637X/810/1/71}, \href
  {https://ui.adsabs.harvard.edu/abs/2015ApJ...810...71F} {810, 71}

\bibitem[\protect\citeauthoryear{{Finkelstein} et~al.,}{{Finkelstein}
  et~al.}{2022}]{finkelstein2022ApJ...940L..55F}
{Finkelstein} S.~L.,  et~al., 2022, \mn@doi [\apjl] {10.3847/2041-8213/ac966e},
  \href {https://ui.adsabs.harvard.edu/abs/2022ApJ...940L..55F} {940, L55}

\bibitem[\protect\citeauthoryear{{Fletcher}, {Tang}, {Robertson}, {Nakajima},
  {Ellis}, {Stark}  \& {Inoue}}{{Fletcher}
  et~al.}{2019}]{Fletcher2019ApJ...878...87F}
{Fletcher} T.~J.,  {Tang} M.,  {Robertson} B.~E.,  {Nakajima} K.,  {Ellis}
  R.~S.,  {Stark} D.~P.,   {Inoue} A.,  2019, \mn@doi [\apj]
  {10.3847/1538-4357/ab2045}, \href
  {https://ui.adsabs.harvard.edu/abs/2019ApJ...878...87F} {878, 87}

\bibitem[\protect\citeauthoryear{{Gaikwad} et~al.,}{{Gaikwad}
  et~al.}{2023}]{Gaikwad2023MNRAS.525.4093G}
{Gaikwad} P.,  et~al., 2023, \mn@doi [\mnras] {10.1093/mnras/stad2566}, \href
  {https://ui.adsabs.harvard.edu/abs/2023MNRAS.525.4093G} {525, 4093}

\bibitem[\protect\citeauthoryear{Greig, Mesinger, Haiman  \& Simcoe}{Greig
  et~al.}{2017}]{Greig2017}
Greig B.,  Mesinger A.,  Haiman Z.,   Simcoe R.~A.,  2017, \mn@doi [\mnras]
  {10.1093/mnras/stw3351}, 466, 4239

\bibitem[\protect\citeauthoryear{Greig, Mesinger  \& Bañados}{Greig
  et~al.}{2019}]{Greig2019}
Greig B.,  Mesinger A.,   Bañados E.,  2019, \mn@doi [\mnras]
  {10.1093/mnras/stz230}, 484, 5094

\bibitem[\protect\citeauthoryear{Greig, Mesinger, Davies, Wang, Yang  \&
  Hennawi}{Greig et~al.}{2022}]{Greig2022}
Greig B.,  Mesinger A.,  Davies F.~B.,  Wang F.,  Yang J.,   Hennawi J.~F.,
  2022, \mn@doi [\mnras] {10.1093/MNRAS/STAC825}, 512, 5390

\bibitem[\protect\citeauthoryear{{Harikane} et~al.,}{{Harikane}
  et~al.}{2023}]{Harikane2023ApJS..265....5H}
{Harikane} Y.,  et~al., 2023, \mn@doi [\apjs] {10.3847/1538-4365/acaaa9}, \href
  {https://ui.adsabs.harvard.edu/abs/2023ApJS..265....5H} {265, 5}

\bibitem[\protect\citeauthoryear{{Inoue} et~al.,}{{Inoue}
  et~al.}{2018}]{Inoue2018PASJ...70...55I}
{Inoue} A.~K.,  et~al., 2018, \mn@doi [\pasj] {10.1093/pasj/psy048}, \href
  {https://ui.adsabs.harvard.edu/abs/2018PASJ...70...55I} {70, 55}

\bibitem[\protect\citeauthoryear{{Ishigaki}, {Kawamata}, {Ouchi}, {Oguri},
  {Shimasaku}  \& {Ono}}{{Ishigaki} et~al.}{2018}]{Ishigaki2018ApJ...854...73I}
{Ishigaki} M.,  {Kawamata} R.,  {Ouchi} M.,  {Oguri} M.,  {Shimasaku} K.,
  {Ono} Y.,  2018, \mn@doi [\apj] {10.3847/1538-4357/aaa544}, \href
  {https://ui.adsabs.harvard.edu/abs/2018ApJ...854...73I} {854, 73}

\bibitem[\protect\citeauthoryear{{Izotov}, {Worseck}, {Schaerer}, {Guseva},
  {Chisholm}, {Thuan}, {Fricke}  \& {Verhamme}}{{Izotov}
  et~al.}{2021}]{Izotov2021MNRAS.503.1734I}
{Izotov} Y.~I.,  {Worseck} G.,  {Schaerer} D.,  {Guseva} N.~G.,  {Chisholm} J.,
   {Thuan} T.~X.,  {Fricke} K.~J.,   {Verhamme} A.,  2021, \mn@doi [\mnras]
  {10.1093/mnras/stab612}, \href
  {https://ui.adsabs.harvard.edu/abs/2021MNRAS.503.1734I} {503, 1734}

\bibitem[\protect\citeauthoryear{{Jin} et~al.,}{{Jin}
  et~al.}{2023}]{Jin2023ApJ...942...59J}
{Jin} X.,  et~al., 2023, \mn@doi [\apj] {10.3847/1538-4357/aca678}, \href
  {https://ui.adsabs.harvard.edu/abs/2023ApJ...942...59J} {942, 59}

\bibitem[\protect\citeauthoryear{Jones et~al.,}{Jones
  et~al.}{2024}]{jones2024jadesmeasuringreionizationproperties}
Jones G.~C.,  et~al., 2024, \mn@doi [arXiv e-prints]
  {10.48550/arXiv.2409.06405}, p. arXiv:2409.06405

\bibitem[\protect\citeauthoryear{{Jung} et~al.,}{{Jung}
  et~al.}{2020}]{Jung2020ApJ...904..144J}
{Jung} I.,  et~al., 2020, \mn@doi [\apj] {10.3847/1538-4357/abbd44}, \href
  {https://ui.adsabs.harvard.edu/abs/2020ApJ...904..144J} {904, 144}

\bibitem[\protect\citeauthoryear{{Kostyuk}, {Nelson}, {Ciardi}, {Glatzle}  \&
  {Pillepich}}{{Kostyuk} et~al.}{2023}]{Kostyuk2023MNRAS.521.3077K}
{Kostyuk} I.,  {Nelson} D.,  {Ciardi} B.,  {Glatzle} M.,   {Pillepich} A.,
  2023, \mn@doi [\mnras] {10.1093/mnras/stad677}, \href
  {https://ui.adsabs.harvard.edu/abs/2023MNRAS.521.3077K} {521, 3077}

\bibitem[\protect\citeauthoryear{{Leonova} et~al.,}{{Leonova}
  et~al.}{2022}]{leonova2022mnras.515.5790l}
{Leonova} E.,  et~al., 2022, \mn@doi [\mnras] {10.1093/mnras/stac1908}, \href
  {https://ui.adsabs.harvard.edu/abs/2022MNRAS.515.5790L} {515, 5790}

\bibitem[\protect\citeauthoryear{{Ma}, {Quataert}, {Wetzel}, {Hopkins},
  {Faucher-Gigu{\`e}re}  \& {Kere{\v{s}}}}{{Ma}
  et~al.}{2020}]{ma2020mnras.498.2001m}
{Ma} X.,  {Quataert} E.,  {Wetzel} A.,  {Hopkins} P.~F.,  {Faucher-Gigu{\`e}re}
  C.-A.,   {Kere{\v{s}}} D.,  2020, \mn@doi [\mnras] {10.1093/mnras/staa2404},
  \href {https://ui.adsabs.harvard.edu/abs/2020MNRAS.498.2001M} {498, 2001}

\bibitem[\protect\citeauthoryear{Mason, Trenti  \& Treu}{Mason
  et~al.}{2015}]{mason2015}
Mason C.~A.,  Trenti M.,   Treu T.,  2015, \mn@doi [\apj]
  {10.1088/0004-637X/813/1/21}, 813, 21

\bibitem[\protect\citeauthoryear{{Mason}, {Treu}, {Dijkstra}, {Mesinger},
  {Trenti}, {Pentericci}, {de Barros}  \& {Vanzella}}{{Mason}
  et~al.}{2018}]{mason2018apj...856....2m}
{Mason} C.~A.,  {Treu} T.,  {Dijkstra} M.,  {Mesinger} A.,  {Trenti} M.,
  {Pentericci} L.,  {de Barros} S.,   {Vanzella} E.,  2018, \mn@doi [\apj]
  {10.3847/1538-4357/aab0a7}, \href
  {https://ui.adsabs.harvard.edu/abs/2018ApJ...856....2M} {856, 2}

\bibitem[\protect\citeauthoryear{{Mason} et~al.,}{{Mason}
  et~al.}{2019}]{Mason2019MNRAS.485.3947M}
{Mason} C.~A.,  et~al., 2019, \mn@doi [\mnras] {10.1093/mnras/stz632}, \href
  {https://ui.adsabs.harvard.edu/abs/2019MNRAS.485.3947M} {485, 3947}

\bibitem[\protect\citeauthoryear{{McGreer}, {Mesinger}  \&
  {D'Odorico}}{{McGreer} et~al.}{2015}]{McGreer2015MNRAS.447..499M}
{McGreer} I.~D.,  {Mesinger} A.,   {D'Odorico} V.,  2015, \mn@doi [\mnras]
  {10.1093/mnras/stu2449}, \href
  {https://ui.adsabs.harvard.edu/abs/2015MNRAS.447..499M} {447, 499}

\bibitem[\protect\citeauthoryear{{Mesinger}, {Furlanetto}  \& {Cen}}{{Mesinger}
  et~al.}{2011}]{Mesinger2011MNRAS.411..955M}
{Mesinger} A.,  {Furlanetto} S.,   {Cen} R.,  2011, \mn@doi [\mnras]
  {10.1111/j.1365-2966.2010.17731.x}, \href
  {https://ui.adsabs.harvard.edu/abs/2011MNRAS.411..955M} {411, 955}

\bibitem[\protect\citeauthoryear{{Miralda-Escud{\'e}}}{{Miralda-Escud{\'e}}}{1998}]{miraldaescide1998apj...501...15m}
{Miralda-Escud{\'e}} J.,  1998, \mn@doi [\apj] {10.1086/305799}, \href
  {https://ui.adsabs.harvard.edu/abs/1998ApJ...501...15M} {501, 15}

\bibitem[\protect\citeauthoryear{{Morales}, {Mason}, {Bruton}, {Gronke},
  {Haardt}  \& {Scarlata}}{{Morales} et~al.}{2021}]{Morales2021ApJ...919..120M}
{Morales} A.~M.,  {Mason} C.~A.,  {Bruton} S.,  {Gronke} M.,  {Haardt} F.,
  {Scarlata} C.,  2021, \mn@doi [\apj] {10.3847/1538-4357/ac1104}, \href
  {https://ui.adsabs.harvard.edu/abs/2021ApJ...919..120M} {919, 120}

\bibitem[\protect\citeauthoryear{{Murray}, {Greig}, {Mesinger}, {Mu{\~n}oz},
  {Qin}, {Park}  \& {Watkinson}}{{Murray}
  et~al.}{2020}]{Murray2020JOSS....5.2582M}
{Murray} S.,  {Greig} B.,  {Mesinger} A.,  {Mu{\~n}oz} J.,  {Qin} Y.,  {Park}
  J.,   {Watkinson} C.,  2020, \mn@doi [The Journal of Open Source Software]
  {10.21105/joss.02582}, \href
  {https://ui.adsabs.harvard.edu/abs/2020JOSS....5.2582M} {5, 2582}

\bibitem[\protect\citeauthoryear{{Mutch} et~al.,}{{Mutch}
  et~al.}{2016}]{Mutch2016MNRAS.463.3556M}
{Mutch} S.~J.,  et~al., 2016, \mn@doi [\mnras] {10.1093/mnras/stw2187}, \href
  {https://ui.adsabs.harvard.edu/abs/2016MNRAS.463.3556M} {463, 3556}

\bibitem[\protect\citeauthoryear{{Mutch}, {Greig}, {Qin}, {Poole}  \&
  {Wyithe}}{{Mutch} et~al.}{2024}]{mutch2024mnras.527.7924m}
{Mutch} S.~J.,  {Greig} B.,  {Qin} Y.,  {Poole} G.~B.,   {Wyithe} J. S.~B.,
  2024, \mn@doi [\mnras] {10.1093/mnras/stad3746}, \href
  {https://ui.adsabs.harvard.edu/abs/2024MNRAS.527.7924M} {527, 7924}

\bibitem[\protect\citeauthoryear{{Naidu}, {Forrest}, {Oesch}, {Tran}  \&
  {Holden}}{{Naidu} et~al.}{2018}]{Naidu2018MNRAS.478..791N}
{Naidu} R.~P.,  {Forrest} B.,  {Oesch} P.~A.,  {Tran} K.-V.~H.,   {Holden}
  B.~P.,  2018, \mn@doi [\mnras] {10.1093/mnras/sty961}, \href
  {https://ui.adsabs.harvard.edu/abs/2018MNRAS.478..791N} {478, 791}

\bibitem[\protect\citeauthoryear{{Naidu} et~al.,}{{Naidu}
  et~al.}{2022}]{Naidu2022ApJ...940L..14N}
{Naidu} R.~P.,  et~al., 2022, \mn@doi [\apjl] {10.3847/2041-8213/ac9b22}, \href
  {https://ui.adsabs.harvard.edu/abs/2022ApJ...940L..14N} {940, L14}

\bibitem[\protect\citeauthoryear{{Nasir} \& {D'Aloisio}}{{Nasir} \&
  {D'Aloisio}}{2020}]{nasir2020mnras.494.3080n}
{Nasir} F.,  {D'Aloisio} A.,  2020, \mn@doi [\mnras] {10.1093/mnras/staa894},
  \href {https://ui.adsabs.harvard.edu/abs/2020MNRAS.494.3080N} {494, 3080}

\bibitem[\protect\citeauthoryear{{Oesch} et~al.,}{{Oesch}
  et~al.}{2016}]{Oesch2016ApJ...819..129O}
{Oesch} P.~A.,  et~al., 2016, \mn@doi [\apj] {10.3847/0004-637X/819/2/129},
  \href {https://ui.adsabs.harvard.edu/abs/2016ApJ...819..129O} {819, 129}

\bibitem[\protect\citeauthoryear{{Ouchi} et~al.,}{{Ouchi}
  et~al.}{2018}]{ouchi2018pasj...70s..13o}
{Ouchi} M.,  et~al., 2018, \mn@doi [\pasj] {10.1093/pasj/psx074}, \href
  {https://ui.adsabs.harvard.edu/abs/2018PASJ...70S..13O} {70, S13}

\bibitem[\protect\citeauthoryear{{Oyarz{\'u}n}, {Blanc}, {Gonz{\'a}lez},
  {Mateo}  \& {Bailey}}{{Oyarz{\'u}n}
  et~al.}{2017}]{Oyarzun2017ApJ...843..133O}
{Oyarz{\'u}n} G.~A.,  {Blanc} G.~A.,  {Gonz{\'a}lez} V.,  {Mateo} M.,
  {Bailey} John~I. I.,  2017, \mn@doi [\apj] {10.3847/1538-4357/aa7552}, \href
  {https://ui.adsabs.harvard.edu/abs/2017ApJ...843..133O} {843, 133}

\bibitem[\protect\citeauthoryear{{Paardekooper}, {Khochfar}  \& {Dalla
  Vecchia}}{{Paardekooper} et~al.}{2015}]{Paardekooper2015MNRAS.451.2544P}
{Paardekooper} J.-P.,  {Khochfar} S.,   {Dalla Vecchia} C.,  2015, \mn@doi
  [\mnras] {10.1093/mnras/stv1114}, \href
  {https://ui.adsabs.harvard.edu/abs/2015MNRAS.451.2544P} {451, 2544}

\bibitem[\protect\citeauthoryear{{Pahl}, {Shapley}, {Steidel}, {Chen}  \&
  {Reddy}}{{Pahl} et~al.}{2021}]{Pahl2021arXiv210402081P}
{Pahl} A.~J.,  {Shapley} A.,  {Steidel} C.~C.,  {Chen} Y.,   {Reddy} N.~A.,
  2021, arXiv e-prints, \href
  {https://ui.adsabs.harvard.edu/abs/2021arXiv210402081P} {p. arXiv:2104.02081}

\bibitem[\protect\citeauthoryear{{Park}, {Mesinger}, {Greig}  \&
  {Gillet}}{{Park} et~al.}{2019}]{Park2019MNRAS.484..933P}
{Park} J.,  {Mesinger} A.,  {Greig} B.,   {Gillet} N.,  2019, \mn@doi [\mnras]
  {10.1093/mnras/stz032}, \href
  {https://ui.adsabs.harvard.edu/abs/2019MNRAS.484..933P} {484, 933}

\bibitem[\protect\citeauthoryear{{P{\'e}rez-Gonz{\'a}lez}
  et~al.,}{{P{\'e}rez-Gonz{\'a}lez} et~al.}{2023}]{PG2023arXiv230202429P}
{P{\'e}rez-Gonz{\'a}lez} P.~G.,  et~al., 2023, \mn@doi [\apjl]
  {10.3847/2041-8213/acd9d0}, \href
  {https://ui.adsabs.harvard.edu/abs/2023ApJ...951L...1P} {951, L1}

\bibitem[\protect\citeauthoryear{{Planck Collaboration} et~al.,}{{Planck
  Collaboration} et~al.}{2020}]{Planck2020A&A...641A...6P}
{Planck Collaboration} et~al., 2020, \mn@doi [\aap]
  {10.1051/0004-6361/201833910}, \href
  {https://ui.adsabs.harvard.edu/abs/2020A&A...641A...6P} {641, A6}

\bibitem[\protect\citeauthoryear{{Qin} et~al.,}{{Qin} et~al.}{2017}]{Qin2017a}
{Qin} Y.,  et~al., 2017, \mn@doi [\mnras] {10.1093/mnras/stx1909}, \href
  {https://ui.adsabs.harvard.edu/abs/2017MNRAS.472.2009Q} {472, 2009}

\bibitem[\protect\citeauthoryear{{Qin}, {Poulin}, {Mesinger}, {Greig}, {Murray}
   \& {Park}}{{Qin} et~al.}{2020}]{qin2020mnras.499..550q}
{Qin} Y.,  {Poulin} V.,  {Mesinger} A.,  {Greig} B.,  {Murray} S.,   {Park} J.,
   2020, \mn@doi [\mnras] {10.1093/mnras/staa2797}, \href
  {https://ui.adsabs.harvard.edu/abs/2020MNRAS.499..550Q} {499, 550}

\bibitem[\protect\citeauthoryear{{Qin}, {Mesinger}, {Bosman}  \& {Viel}}{{Qin}
  et~al.}{2021}]{qin2021mnras.506.2390q}
{Qin} Y.,  {Mesinger} A.,  {Bosman} S. E.~I.,   {Viel} M.,  2021, \mn@doi
  [\mnras] {10.1093/mnras/stab1833}, \href
  {https://ui.adsabs.harvard.edu/abs/2021MNRAS.506.2390Q} {506, 2390}

\bibitem[\protect\citeauthoryear{{Qin}, {Wyithe}, {Oesch}, {Illingworth},
  {Leonova}, {Mutch}  \& {Naidu}}{{Qin} et~al.}{2022}]{qin2022mnras.510.3858q}
{Qin} Y.,  {Wyithe} J. S.~B.,  {Oesch} P.~A.,  {Illingworth} G.~D.,  {Leonova}
  E.,  {Mutch} S.~J.,   {Naidu} R.~P.,  2022, \mn@doi [\mnras]
  {10.1093/mnras/stab3733}, \href
  {https://ui.adsabs.harvard.edu/abs/2022MNRAS.510.3858Q} {510, 3858}

\bibitem[\protect\citeauthoryear{{Qin}, {Balu}  \& {Wyithe}}{{Qin}
  et~al.}{2023}]{Qin2023MNRAS.526.1324Q}
{Qin} Y.,  {Balu} S.,   {Wyithe} J. S.~B.,  2023, \mn@doi [\mnras]
  {10.1093/mnras/stad2448}, \href
  {https://ui.adsabs.harvard.edu/abs/2023MNRAS.526.1324Q} {526, 1324}

\bibitem[\protect\citeauthoryear{{Schaerer}}{{Schaerer}}{2003}]{Schaerer2003A&A...397..527S}
{Schaerer} D.,  2003, \mn@doi [\aap] {10.1051/0004-6361:20021525}, \href
  {https://ui.adsabs.harvard.edu/abs/2003A&A...397..527S} {397, 527}

\bibitem[\protect\citeauthoryear{{Sobral} et~al.,}{{Sobral}
  et~al.}{2018}]{Sobral2018MNRAS.477.2817S}
{Sobral} D.,  et~al., 2018, \mn@doi [\mnras] {10.1093/mnras/sty782}, \href
  {https://ui.adsabs.harvard.edu/abs/2018MNRAS.477.2817S} {477, 2817}

\bibitem[\protect\citeauthoryear{{Stark} et~al.,}{{Stark}
  et~al.}{2015}]{Stark2015MNRAS.454.1393S}
{Stark} D.~P.,  et~al., 2015, \mn@doi [\mnras] {10.1093/mnras/stv1907}, \href
  {https://ui.adsabs.harvard.edu/abs/2015MNRAS.454.1393S} {454, 1393}

\bibitem[\protect\citeauthoryear{{Stark} et~al.,}{{Stark}
  et~al.}{2017}]{Stark2017MNRAS.464..469S}
{Stark} D.~P.,  et~al., 2017, \mn@doi [\mnras] {10.1093/mnras/stw2233}, \href
  {https://ui.adsabs.harvard.edu/abs/2017MNRAS.464..469S} {464, 469}

\bibitem[\protect\citeauthoryear{{Steidel}, {Bogosavljevi{\'c}}, {Shapley},
  {Reddy}, {Rudie}, {Pettini}, {Trainor}  \& {Strom}}{{Steidel}
  et~al.}{2018}]{Steidel2018ApJ...869..123S}
{Steidel} C.~C.,  {Bogosavljevi{\'c}} M.,  {Shapley} A.~E.,  {Reddy} N.~A.,
  {Rudie} G.~C.,  {Pettini} M.,  {Trainor} R.~F.,   {Strom} A.~L.,  2018,
  \mn@doi [\apj] {10.3847/1538-4357/aaed28}, \href
  {https://ui.adsabs.harvard.edu/abs/2018ApJ...869..123S} {869, 123}

\bibitem[\protect\citeauthoryear{{Tang}, {Stark}, {Topping}, {Mason}  \&
  {Ellis}}{{Tang} et~al.}{2024}]{Tang2024arXiv240801507T}
{Tang} M.,  {Stark} D.~P.,  {Topping} M.~W.,  {Mason} C.,   {Ellis} R.~S.,
  2024, \mn@doi [arXiv e-prints] {10.48550/arXiv.2408.01507}, \href
  {https://ui.adsabs.harvard.edu/abs/2024arXiv240801507T} {p. arXiv:2408.01507}

\bibitem[\protect\citeauthoryear{{Wang} et~al.,}{{Wang}
  et~al.}{2020}]{Wang2020ApJ...896...23W}
{Wang} F.,  et~al., 2020, \mn@doi [\apj] {10.3847/1538-4357/ab8c45}, \href
  {https://ui.adsabs.harvard.edu/abs/2020ApJ...896...23W} {896, 23}

\bibitem[\protect\citeauthoryear{{Whitler}, {Mason}, {Ren}, {Dijkstra},
  {Mesinger}, {Pentericci}, {Trenti}  \& {Treu}}{{Whitler}
  et~al.}{2020}]{Whitler2020MNRAS.495.3602W}
{Whitler} L.~R.,  {Mason} C.~A.,  {Ren} K.,  {Dijkstra} M.,  {Mesinger} A.,
  {Pentericci} L.,  {Trenti} M.,   {Treu} T.,  2020, \mn@doi [\mnras]
  {10.1093/mnras/staa1178}, \href
  {https://ui.adsabs.harvard.edu/abs/2020MNRAS.495.3602W} {495, 3602}

\bibitem[\protect\citeauthoryear{{Willott} et~al.,}{{Willott}
  et~al.}{2024}]{Willott2024ApJ...966...74W}
{Willott} C.~J.,  et~al., 2024, \mn@doi [\apj] {10.3847/1538-4357/ad35bc},
  \href {https://ui.adsabs.harvard.edu/abs/2024ApJ...966...74W} {966, 74}

\bibitem[\protect\citeauthoryear{{Witstok} et~al.,}{{Witstok}
  et~al.}{2024a}]{Witstok2024arXiv240816608W}
{Witstok} J.,  et~al., 2024a, \mn@doi [arXiv e-prints]
  {10.48550/arXiv.2408.16608}, \href
  {https://ui.adsabs.harvard.edu/abs/2024arXiv240816608W} {p. arXiv:2408.16608}

\bibitem[\protect\citeauthoryear{{Witstok} et~al.,}{{Witstok}
  et~al.}{2024b}]{Witstok2024A&A...682A..40W}
{Witstok} J.,  et~al., 2024b, \mn@doi [\aap] {10.1051/0004-6361/202347176},
  \href {https://ui.adsabs.harvard.edu/abs/2024A&A...682A..40W} {682, A40}

\bibitem[\protect\citeauthoryear{{Witten} et~al.,}{{Witten}
  et~al.}{2024}]{Witten2024NatAs...8..384W}
{Witten} C.,  et~al., 2024, \mn@doi [Nature Astronomy]
  {10.1038/s41550-023-02179-3}, \href
  {https://ui.adsabs.harvard.edu/abs/2024NatAs...8..384W} {8, 384}

\bibitem[\protect\citeauthoryear{{Wold} et~al.,}{{Wold}
  et~al.}{2022}]{wold2022apj...927...36w}
{Wold} I. G.~B.,  et~al., 2022, \mn@doi [\apj] {10.3847/1538-4357/ac4997},
  \href {https://ui.adsabs.harvard.edu/abs/2022ApJ...927...36W} {927, 36}

\bibitem[\protect\citeauthoryear{{Xu}, {Wise}, {Norman}, {Ahn}  \&
  {O'Shea}}{{Xu} et~al.}{2016}]{Xu2016ApJ...833...84X}
{Xu} H.,  {Wise} J.~H.,  {Norman} M.~L.,  {Ahn} K.,   {O'Shea} B.~W.,  2016,
  \mn@doi [\apj] {10.3847/1538-4357/833/1/84}, \href
  {http://adsabs.harvard.edu/abs/2016ApJ...833...84X} {833, 84}

\bibitem[\protect\citeauthoryear{{Zhu} et~al.,}{{Zhu}
  et~al.}{2024}]{Zhu2024MNRAS.533L..49Z}
{Zhu} Y.,  et~al., 2024, \mn@doi [\mnras] {10.1093/mnrasl/slae061}, \href
  {https://ui.adsabs.harvard.edu/abs/2024MNRAS.533L..49Z} {533, L49}

\makeatother
\end{thebibliography}

% Don't change these lines
\bsp	% typesetting comment
\label{lastpage}
\end{document}